# Selection Induced Contrast Estimate (SICE) Effect: An Attempt to Quantify the Impact of Some Patient Selection Criteria in Randomized Clinical Trials

Junshui Ma* and Daniel J. Holder

Merck & Co., Inc., Kenilworth, NJ, USA

## ABSTRACT

Defining the Inclusion/Exclusion (I/E) criteria of a trial is one of the most important steps during a trial design. Increasingly complex I/E criteria potentially create information imbalance and transparency issues between the people who design and run the trials and those who consume the information produced by the trials. In order to better understand and quantify the impact of a category of I/E criteria on observed treatment effects, a concept, named the Selection Induced Contrast Estimate (SICE) effect, is introduced and formulated in this paper. The SICE effect can exist in controlled clinical trials when treatment affects the correlation between a marker used for selection and the response of interest. This effect is demonstrated with both simulations and real clinical trial data. Although the statistical elements behind the SICE effect have been well studied, explicitly formulating and studying this effect can benefit several areas, including better transparency in I/E criteria, meta-analysis of multiple clinical trials, treatment effect interpretation in real-world medical practice, etc.

*Corresponding author, email: Junshui_ma@merck.com

# 1. INTRODUCTION

Randomized Clinical Trials (RCTs) are considered the gold standard for the identification of treatment effect. RCTs always have a set of patient selection criteria, i.e. Inclusion/Exclusion (I/E) criteria, to unambiguously define their targeted populations. The identified treatment effects are only true for the targeted population (Food and Drug Administration, 2012). Since the I/E criteria are often critical for the success of a trial, defining them becomes one of the most important steps when designing a trial. I/E criteria also become increasingly complex.

Complex I/E criteria potentially create information imbalance and transparency issues between the people who design and run the trials (e.g. pharmaceutical companies) and the people who consume the information produced by the trials (e.g. physicians and patients). Although it is a general practice to provide I/E criteria along with study results, people outside sometimes lack the insights to understand, among the dozens of I/E criteria, which ones are more critical to the trial results and should be followed more strictly than the others. For example, in clinical trials to study insomnia drugs, a group of I/E criteria are specified to ensure that the volunteers are indeed insomnia patients. Latency to Persistent Sleep (LPS) is to measure of how long it takes for a volunteer to get to stable sleep stage. Some trials choose LPS >= 15 minutes as an inclusion criterion, while the others choose LPS >= 20 minutes. Not everyone knows that this subtle difference can cause large differences in the resultant treatment effect size.

Lacking the insights into the impact of each I/E criterion can cause serious issues in various contexts. It can play a big role in the discrepancy between treatment effects reported from RCTs and those observed in real-world data. It also can invalidate research efforts that attempt to borrow information from historical trials.

Solutions to this complex issue require the collaboration among many parties, including regulatory agencies, pharmaceutical companies, clinicians, and statisticians. This paper is an effort to gain insights from a statistical perspective into a category of impactful I/E criteria.

Before diving into the technical details, a simulated 2-arm (i.e. drug vs. control) parallel trial is used to illustrate the general idea (see Appendix A for the simulation code). Assume that the trial recruits 1000 patients for each treatment arm. Let a random variable (r.v.) $x$ represent a patient's baseline measurement, while a r.v. $y$ represents the post-treatment clinical endpoint. Assume that $x$ and $y$ follow a bivariate normal distribution. If the patient is finally randomized to the control arm, the distribution of these two variables is represented in (1a); if the patient is randomized to the drug arm, the distribution is represented in (1b). Note that $y_c$ denotes the clinical endpoints for patients in the control arm, while $y_t$ for the drug arm. (1a) and (1b) show neither treatment changes the population mean, which remains 0, and they only slightly alter the variance and correlation of the clinical endpoints.

$$\begin{bmatrix} x \\ y_c \end{bmatrix} \sim N(\begin{bmatrix} 0 \\ 0 \end{bmatrix}, \begin{bmatrix} 1 & 0.2 \\ 0.2 & 1 \end{bmatrix}) \tag{1a}$$

$$\begin{bmatrix} x \\ y_t \end{bmatrix} \sim N(\begin{bmatrix} 0 \\ 0 \end{bmatrix}, \begin{bmatrix} 1 & 0.3 \times 1.2 \\ 0.3 \times 1.2 & (1.2)^2 \end{bmatrix}) \tag{1b}$$

Since the post-treatment mean difference between $y_t$ and $y_c$ is 0, the probability for any reasonable statistical procedure, e.g. 2-sample t-test or ANCOVA, to claim a significant drug effect should be controlled at a prespecified type-1 error rate, normally 5%.

Now, we introduce a patient selection process (i.e. an inclusion criterion). That is, only the patients with $x \geq 0.6$ are randomized to one of the two treatment arms. Then, any reasonable statistical procedure will almost surely (with a probability of about 98%) claim a significant drug effect. The "discrepancy" between the claimed drug effects with and without the patient selection (i.e. $x \geq 0.6$) arises because the drug effects are regarding different targeted populations. That is, the I/E criterion of $x \geq 0.6$ is an impactful criterion and deserves special attention.

The difference in treatment effects between two populations created by different patient selection criteria is the focus of this paper. We name this difference as *Selection Induced Contrast Estimate (SICE) effect*, which is subsequently formulated and quantified.

Explicitly defining this concept and bringing it to the attention of the statistical and medical communities have several benefits. First, some therapeutic areas, e.g. neuroscience, are notorious for its unpredictable failures in late-stage clinical studies. The drug effect observed in earlier-stage clinical studies suddenly "disappears" in later clinical trials. Some of these surprises can be explained with the SICE effect.

Second, as previous mentioned, a better understanding of the SICE effect provides us with insights into the impacts of some I/E criteria. The SICE effect allows us to foresee which I/E criteria can be more impactful than the others. This knowledge is useful in many applications. For example, meta-data analyses often pool results from multiple studies. The SICE effect can help us gain insights into selecting the proper studies to pool for meta-analysis.

In addition, the SICE effect can help to quantify the potential discrepancy between the drug effects observed in a clinical trial and those observed in the real-world setting. Although a drug label provides detailed information of the drug's targeted population, prescribers may consider expanding use to patients they consider similar. Diseases are often defined broadly in medical practice, while clinical trials for drug approval need to select patients base on unambiguous I/E criteria. For example, insomnia is medically defined as "a sleep disorder in which there is an inability to fall asleep or to stay asleep as long as desired" (Roth, 2007). In contrast, a trial to test an insomnia drug requires specifying an unambiguous insomniac criterion, such as "mean latency to persistent sleep (LPS) on both baseline nights of $\geq$ 20 minutes" (Eisai Inc., 2013). If this drug is approved for treating insomnia, it is generally acceptable for a doctor to prescribe the drug without requiring the patient to spend two nights at a sleep center to ensure his/her mean LPS of $\geq$ 20 minutes. As a result, the population to which doctors prescribe the drug can be different from the one tested in the clinical trials. The concept of the SICE effect can help some medical communities understand potentially how much the treatment effect observed in practice may differ from that reported in the clinical trials.

The SICE effect is related to, but different from, some other well studied effects induced by a selection process, e.g. selection bias and the Heckman correction (Heckman, 1979), Regression towards the Mean (RTM) effect (Galton, 1886; Stigler 1997; Barnett, van der Pols, and Dobson, 2005), etc. The key difference is that these selection-induced effects focus the impact of selection on one population, while the SICE effect is about the impact on the mean difference between two populations. In fact, an alternative interpretation of the SICE effect is as the difference between the RTM effect between the control and the treated population. Both the control and the drug groups have a selection procedure, which induces two separate RTM effects in the two groups. However, the sizes of the RTM effects can be different in these two groups. The SICE effect is indeed the difference between the RTM effects of the two groups. Detailed explanation is provided in Subsection 2.1. This alternative perspective highlights the issue that although many (Senn and Brown, 1985; Barnett, van der Pols, and Dobson, 2005) suggest that the RTM effect is completely corrected by the RTM of the control group, this may not always be the case. Although, compared with the RTM effect, the SICE effect is relatively small in magnitude, it can become statistically significant in large-scale confirmatory clinical trials, as will be demonstrated with both simulation results in Subsection 2.2 and a clinical trial dataset in Section 4.

The remainder of this paper is organized as follows. After the SICE effect is formulated and discussed in Subsection 2.1, simulations are presented to illustrate how the magnitude of the SICE effect is impacted by different study scenarios. The goal of Section 3 is primarily to show that, given a selected population, it is hard to correct the SICE effect even under a very strong assumption about the data. The materials in this section are not crucial to understand the concept of SICE effect. In Section 4, a real clinical trial dataset is used to demonstrate that, due to the SICE effect, the impact of different patient selection criteria on the magnitude of the observed treatment effect cannot be ignored in larger clinical trials. A summary and more discussions are provided in Section 5.

## 2. THE SELECTION INDUCED CONTRAST ESTIMATE (SICE) EFFECT

2.1. Formulating the SICE effect

Let us assume that the joint distributions of the selection measurement, $\mathbf{x}$, and the clinical endpoint, $y$, of patients being potentially assigned to the control and the treatment groups respectively follow the two distributions in (2).

$$\begin{bmatrix} \mathbf{x} \\ y_c \end{bmatrix} \sim P_c(\boldsymbol{\theta}_c) \tag{2a}$$

$$\begin{bmatrix} \mathbf{x} \\ y_t \end{bmatrix} \sim P_t(\boldsymbol{\theta}_t), \tag{2b}$$

where $P_c(\boldsymbol{\theta}_c)$ is the distribution of the patients in the control group with a set of distribution parameters $\boldsymbol{\theta}_c$, while $P_t(\boldsymbol{\theta}_t)$ is that of the patients in the treatment group.

For simpler notation, we subsequently define the two patient populations as pre-selection and post-selection populations. It is easy to see all the concepts and results are readily applicable to two selected populations obtained by different selection criteria.

Define the treatment effect over the pre-selection population (i.e. the larger population), $e_o$, as

$$e_o = \mathbf{E}[y_t] - \mathbf{E}[y_c], \text{ and} \tag{3}$$

the treatment effect over the selected sub-population, $e_s$, as

$$e_s = \mathbf{E}[y_t \mid \mathbf{x} \in \Omega] - \mathbf{E}[y_c \mid \mathbf{x} \in \Omega], \tag{4}$$

where $\Omega$ represents a set of selection criteria. The SICE effect, $E_{SIC}$, is thus defined as

$$E_{SICE} \triangleq e_o - e_s. \tag{5}$$

Thus, if $\mathbf{x}$ is not independent of both $y_t$ and $y_c$, $E_{SICE}$ can be non-zero.

To simplify the subsequent presentation, we focus on one selection criterion at a time, and assume that all measurements follow a normal distribution. Equation (2) thus becomes two bivariate normal distributions, represented by Equation (6).

$$\begin{bmatrix} x \\ y_c \end{bmatrix} \sim N(\begin{bmatrix} \mu \\ \mu_c \end{bmatrix}, \begin{bmatrix} \sigma^2 & \rho_c \sigma_c \sigma \\ \rho_c \sigma_c \sigma & \sigma_c^2 \end{bmatrix}) \tag{6a}$$

$$\begin{bmatrix} x \\ y_t \end{bmatrix} \sim N(\begin{bmatrix} \mu \\ \mu_t \end{bmatrix}, \begin{bmatrix} \sigma^2 & \rho_t \sigma_t \sigma \\ \rho_t \sigma_t \sigma & \sigma_t^2 \end{bmatrix}) \tag{6b}$$

The remainder of this section assumes that the selection criterion has a form of $x > a$, where $a$ is a selection threshold. The derivation of the results in this section is provided in Appendix B.

Results with other forms of selection criteria, such as $x < a$, or $b > x > a$, can be obtained in a similar way.

Under these assumptions, the SICE effect, $E_{SICE}$, in (5) becomes

$$E_{SICE} = S(z)[\rho_c \sigma_c - \rho_t \sigma_t] = S(z)[cov(x, y_c) - cov(x, y_t)]/\sigma, \tag{7}$$

where $cov(x, y)$ is the covariance between random variables $x$ and $y$; $z = (a - \mu)/\sigma$, where $\mu$ and $\sigma$ are defined in (6) as the mean and STD of the baseline measure $x$.

$S(z) = \phi(z) / \Phi(-z)$, where $\phi(z) = (\sqrt{2\pi})^{-1} \exp(-z^2/2)$ and $\Phi(z) = \int_{-\infty}^{z} \phi(t) dt$.

Equation (7) helps us better understand the SICE effect. First, since $S(z) > 0$, the SICE effect, $E_{SICE}$, is non-zero whenever $cov(x, y_t) \neq cov(x, y_c)$. Second, the difference between $cov(x, y_t)$ and $cov(x, y_c)$ decides whether the observed treatment effect in the selected population is inflated or deflated by the SICE effect. Third, the SICE effect disappears, if the selection measurement x is uncorrelated with the clinical endpoint y. Finally, the fact that $S(z)$ is a monotonically increasing function suggests that the more selective the selection criterion is (i.e. the threshold $a$ is bigger), the more prominent the SICE effect is.

If $y'_c$ and $y'_t$ represent the responses of the selected patients treated with control or drug respectively, and the treatment effect, $e_s$, in Equation (4) is estimated using sample means, i.e.

$$\hat{e}_s = \bar{y}'_t - \bar{y}'_c, \tag{8}$$

where $\bar{y}'_g = \sum_{i=1}^{N_g} y'_{g,i} / N_g$, $g \in \{t, c\}$, $N_c$ and $N_t$ are the number of patients recruited in the control and the drug groups, respectively, after the selection process.

It can be shown that $\hat{e}_s$ follows a normal distribution with mean, $\mathbf{E}[\hat{e}_s]$, and variance, $\mathbf{V}[\hat{e}_s]$, shown in (9).

$$\mathbf{E}[\hat{e}_s] = e_s = e_o - E_{SICE} \qquad (9a)$$

$$\mathbf{V}[\hat{e}_s] = (\frac{\sigma_t^2}{N_t} + \frac{\sigma_c^2}{N_c}) - S(z)(S(z)-z)(\frac{\rho_t^2 \sigma_t^2}{N_t} + \frac{\rho_c^2 \sigma_c^2}{N_c}) \qquad (9b)$$

Equations (3)-(5), (7) and (9) suggest that the SICE effect does not depend on the number of patients, while the variance of its estimate decreases with more patients. Thus, there is always a chance for a non-zero SICE effect to be statistically significant when the size of the clinical trial becomes sufficiently large.

Equation (9) suggests that a patient selection process has an unpredictable consequence on the magnitude of the treatment effect, if the relationship between the baseline measurement $x$ and clinical endpoint $y$ is unknown. In a clinical trial similar to our setting, if the treatment effect over the pre-selection patient population, $e_o$, is fixed, the estimated treatment effect over the selected patient population, $\hat{e}_s$, can be larger or smaller than $e_o$ due to the SICE effect. According to (9a), the estimated effect, $\hat{e}_s$, is less than $e_o$, if $E_{SICE} > 0$, and is larger than $e_o$ if $E_{SICE} < 0$. Whether the SICE effect, $E_{SICE}$, is positive or negative is decided by the sign of $[cov(x, y_c) - cov(x, y_t)]$, which is often unknown before the trial. However, in some cases, the relationship between $x$ and $y$ can be known. For example, if we know $x$ is a biomarker that is positively correlated with the drug mechanism (e.g. PD-L1 biomarker regarding anti-PD-L1 treatment), but uncorrelated with the control, we will get $E_{SICE} < 0$, and a larger treatment effect in the selected population.

Secondly, a patient selection process does help reducing the variance of the estimated effect. Let us assume that the data of patients before and after a selection process are available, and the treatment effects over pre-selection and post-select populations can be estimated using sample means as $\hat{e}_o$ and $\hat{e}_s$. Equation (8) and (9b) suggests that the variance of the estimate, $\hat{e}_s$, is less than that of $\hat{e}_o$. That is, $\mathbf{V}[\hat{e}_s] < (\frac{\sigma_t^2}{N_t} + \frac{\sigma_c^2}{N_c}) = \mathbf{V}[\hat{e}_o]$, since $S(z) > z$ and $S(z) > 0$. Therefore, if the number of patients in each treatment group is fixed (e.g. pre-specified in a trial protocol), a patient selection process will reduce the variance of the estimator, and thus increases the statistical power of detecting a treatment effect.

As mentioned in the Introduction, the SICE effect can also be explained using the concept of the RTM effect. The RTM effects for the control and the treatment groups can be represented as

$$\frac{\mathbf{E}[y_c | x > a] - \mu_c}{\sigma_c} = \rho_c \frac{\mathbf{E}[x | x > a] - \mu}{\sigma} \tag{10a}$$

$$\frac{\mathbf{E}[y_t | x > a] - \mu_t}{\sigma_t} = \rho_t \frac{\mathbf{E}[x | x > a] - \mu}{\sigma} \tag{10b}$$

If $\rho_c \sigma_c \neq \rho_t \sigma_t$, the RTM effect of $\mathbf{E}[x | x > a]$ of the control group will not cancel out that of the treatment group. What is left is exactly the SICE effect, according to (7).

It is possible to derive similar results for other parametric distributions, such as the central t-distributions for measurements with a heavy-tail distribution and the non-central t-distributions for measurements with an asymmetric distribution. Because the resultant formulas are complex, they are not provided in this paper. Instead, simulation results in Subsection 2.2 are used to demonstrate that the SICE effect under a bivariate t-distribution has a similar form as that in (7).

2.2. Simulation

First, we study the impacts of data distributions and selection thresholds on the SICE effect. The settings of the parameters of the simulation are listed in Table 1.

Table 1

| Distribution | $\mu_c = \mu$ | $\mu_t$ | $\sigma_c = \sigma$ | $\sigma_t$ | $\rho_c$ | $\rho_t$ | $(a-\mu)/\sigma$ | $N_g$ |
|---|---|---|---|---|---|---|---|---|
| $t(df=3)$, $t(df=8)$, normal | 0 | 0 | 1.00 | 0.50, 0.85, 1.00, 1.15, 1.50 | 0.2 | 0.1, 0.2, 0.3, 0.6 | q(0.25), q(0.5), q(0.75) | 1500 |

In Table 1, $t(df=n)$ denotes a t-distribution with degrees of freedom of $n$, $n=3$ or 8. $q(p)$ denotes the quantile of a probability, p, under the corresponding data distribution, and p = 0.25, 0.5, or 0.75. It is used to determine the select threshold $a$. For example, when p=0.25, the selection threshold $a$ should be set so that 25% patients would be excluded by the selection procedure. $N_g$ is the number of patients in the control or the treatment group. $\mu_c = \mu$ and $\sigma_c = \sigma$ ensure that the selection measurement and the clinical endpoint at the control group follow two identical distributions.

The parameters in Table 1 produce 180 different parameter combinations, i.e. simulation scenarios. For each simulation scenario, 2000 trials were simulated. According to (3), (4), (5), (8), and (9a), the estimate of the SICE effect, $\hat{E}_{SICE}$, is a function of the parameters in Table 1, and is calculated as

$$\hat{E}_{SICE} = \sum_{i=1}^{2000} \hat{E}_{SICE}^{(i)} / 2000 = \sum_{i=1}^{2000} \hat{e}_s^{(i)} / 2000 - (\mu_t - \mu_c),$$

where $\hat{e}_s^{(i)}$ and $\hat{E}_{SICE}^{(i)}$ are, respectively, the estimate of the treatment effect for the selected patients and that of the corresponding SICE effect in the $i^{th}$ trial out of the 2000 trials. $\hat{E}_{SICE}^{(i)}$ asymptotically follows a normal distribution and is claimed to be significant if a null hypothesis of $\hat{E}_{SICE}^{(i)} = 0$ is excluded with a significance level of 0.05. The *probability of* $\hat{E}_{SICE}$ *being significant* can be calculated as the ratio of the number of simulated trials, whose estimate, $\hat{E}_{SICE}^{(i)}$, is statistically significant, to the total number of simulated trials, i.e. 2000 in this case. This measurement can also be considered as the type-1 error of the treatment effect over the pre-selection population, $e_o$, caused by the SICE effect.

Estimate $\hat{e}_s^{(i)}$ was obtained using both the 2-sample t-test and the ANCOVA method. Both methods produced almost the same results, and the results of the ANCOVA method are just slightly less variable. Thus, only the results of the ANCOVA method are reported in Figure 1 to keep it concise.

The left panel of the Figure 1 suggests that the SICE effect under a t-distribution should have a similar form as (7), and can be represented as $f(a, df)[\rho_c \sigma_c - \rho_t \sigma_t]$, where $f(a, df)$ denotes a function increasing with the selection threshold, $a$, but decreasing with degrees of freedom, $df$, of the t-distribution. That is, both stricter selection criteria and heavier tail in the data distribution make the SICE effect larger. The right panel of Figure 1 shows that, with a high probability, the SICE effect can cause that a treatment effect is declared in the selected population when the true difference between the means in the pre-selection population is zero.

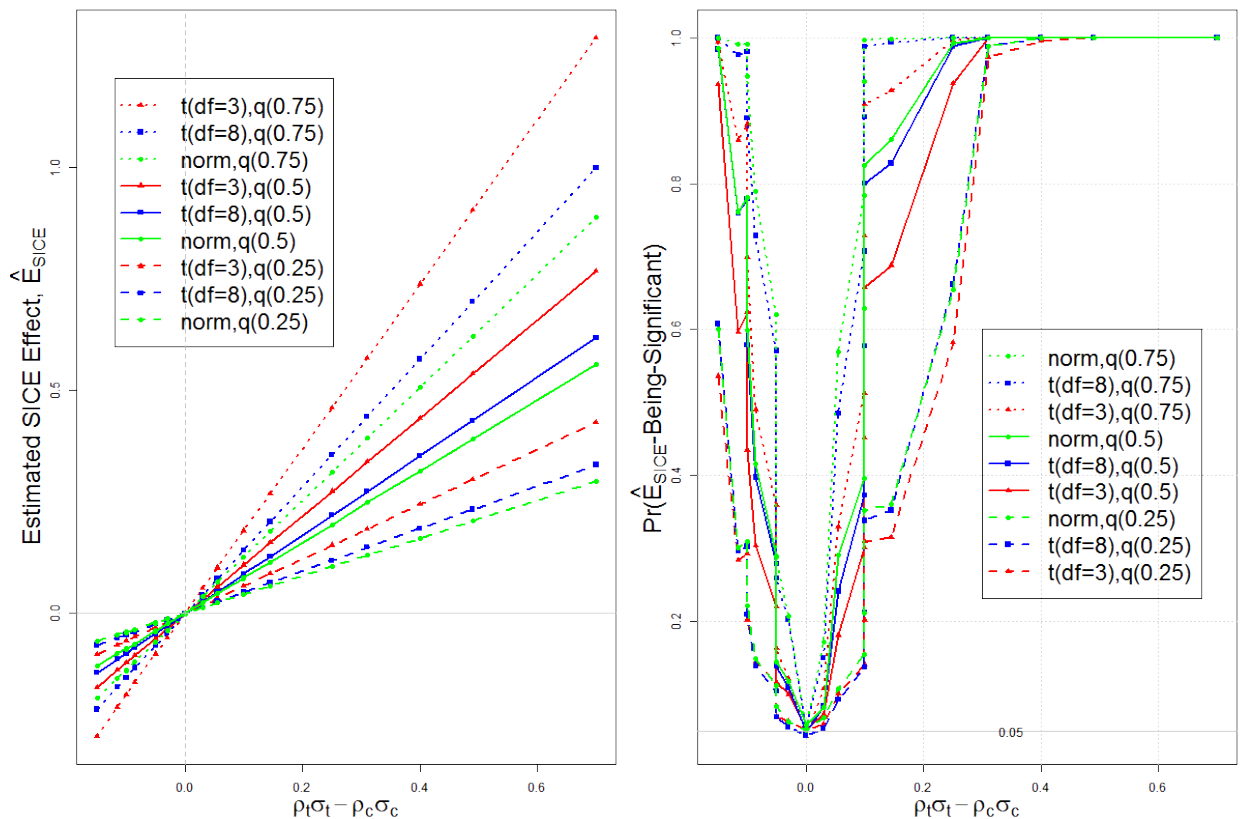

Figure 1 The SICE effect under different data distributions and selection thresholds

Next, we study the impact of the number of patients in each treatment group on the estimated treatment effect induced by the SICE effect. The simulation settings are specified in Table 2, and their results are in Figure 2. Note that the magnitude of the estimated SICE effect is the same as that of the estimated treatment effect in this setting, since $e_o = 0$.

Table 2

| Distribution | $\mu_c = \mu$ | $\mu_t$ | $\sigma_c = \sigma$ | $\sigma_t$ | $\rho_c$ | $\rho_t$ | $(a-\mu)/\sigma$ | $N_g$ |
|---|---|---|---|---|---|---|---|---|
| *normal* | 0.0 | 0.0 | 1.00 | 0.50, 0.85, 1.00, 1.15, 1.50 | 0.2 | 0.1, 0.2, 0.3, 0.6 | q(0.5) | 20, 50, 150, 500, 1000, 2000, 5000 |

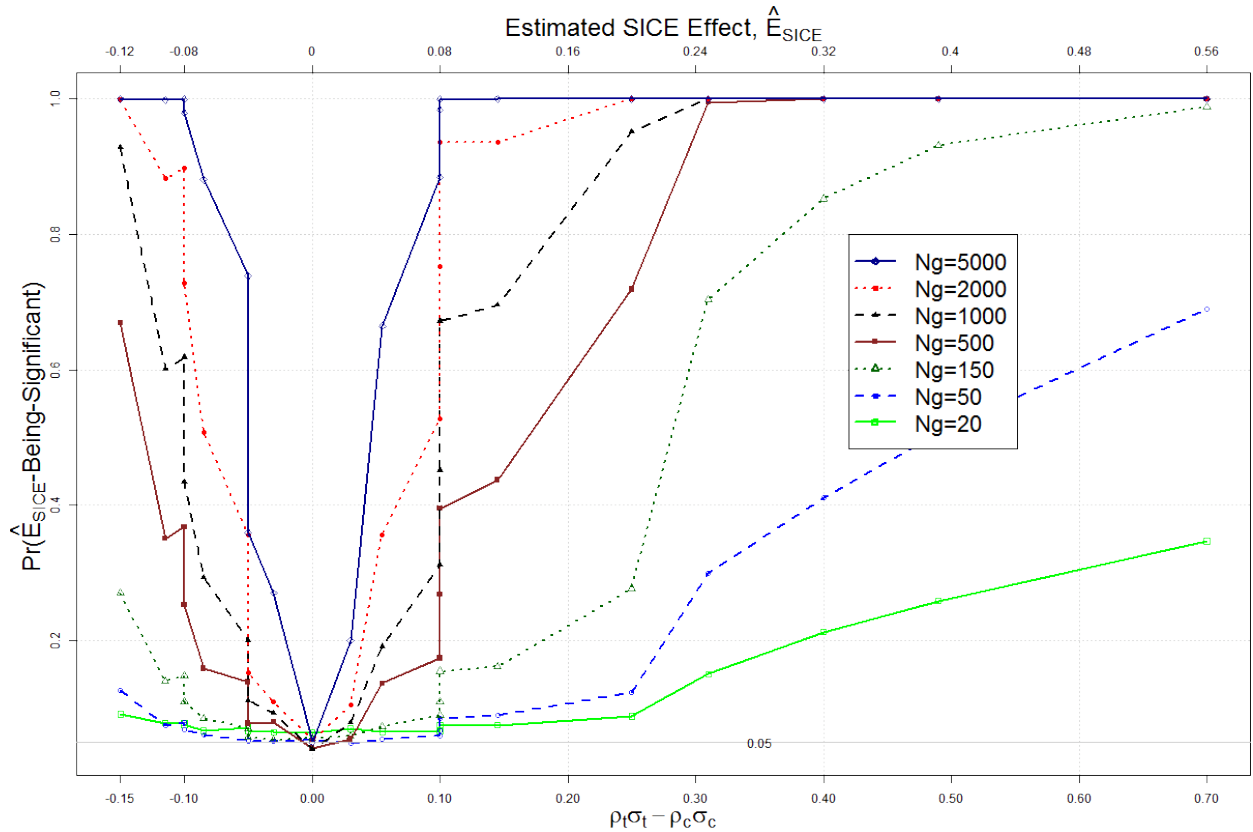

Figure 2 The SICE effect under different numbers of patients

According to (7), the SICE effect, $E_{SICE}$, is a linear function of $\rho_c\sigma_c - \rho_t\sigma_t$, and is not a function of number of patients. Thus, the value of the estimated SICE effect, $\hat{E}_{SICE}$, is plotted at the upper x-axis.

The simulation results suggest that, when the size of clinical trials becomes larger, the probability for the SICE effect to be significant becomes increasingly pronounced, even for the cases where $\rho_t\sigma_t - \rho_c\sigma_c$ is small.

Note that the curves of $\Pr(\hat{E}_{SICE} - Being - Significant)$ in Figure 1 and 2 all have some unsmooth jumps at $\rho_t\sigma_t - \rho_c\sigma_c = 0.1$, -0.05, and -0.1. Let's take $\rho_t\sigma_t - \rho_c\sigma_c = 0.1$ as an example to explain the jumps. More than one parameter combinations in Table 1 and 2, e.g $\rho_t = 0.2, \sigma_t = 1.5$ and

$\rho_t = 0.3, \sigma_t = 1$, can produce $\rho_t\sigma_t - \rho_c\sigma_c = 0.1$. Although the estimated SICE effect, $\hat{E}_{SICE}$, in both cases are the same, the probability for $\hat{E}_{SICE}$ to be significant is smaller for the cases where $\sigma_t$ is larger, because the variance of $\hat{E}_{SICE}$ increases with $\sigma_t$ according to (9b). This causes the jumps in the curves in both figures.

## 3. A METHOD TO ESTIMATE THE TREATMENT EFFECT IN THE PRE-SELECTION POPULATION

This section attempts to address a question: given a selected population and some strong assumptions on the data, is it possible to mitigate the SICE effect? As previously mentioned, materials in this section are not crucial for understanding the concept of SICE effect.

3.1. A Maximum Likelihood method to estimate the treatment effect in the pre-selection population

The SICE effect quantifies the discrepancy in treatment effects observed over the pre-selection population and the post-selection population. In some cases, the ability to estimate the treatment effect in the pre-selection population, $e_o$, using the available post-selection population becomes desirable. For example, if the treatment effect, $e_o$, is estimated to be significant, the population targeted by the treatment may be expanded by removing a selective criterion. Also, the estimate of $e_o$ can provide a quantitative measure on how likely the drug can be effective regarding the pre-selection population.

The data distribution of the pre-selection population should be assumed to follow distributions with a finite number of parameters to make it possible to estimate $e_o$ only with the data of the selected patients. Again, a bivariate normal distribution is assumed, because abundant mathematical tools are available in this context (Senn and Brown, 1985; James, 1973; Cohen, 1955; Cohen, 1957). Senn and Brown (1985) suggest that the maximum likelihood (ML) estimator proposed in Cohen (1955), which we name as "Cohen ML" for short, is a better method than the method of moment approach proposed in James (1973). Thus, we subsequently focus on the Cohen ML estimator. In order to make this paper self-contained, some equations in Cohen (1955) are duplicated in this section.

A number of notation conventions are used. $x$, $y_t$ and $y_c$ denote measurements in the pre-selection population, while $x'_c$, $x'_t$, $y'_c$ and $y'_t$ denote the measurements of the selected patients in the control or the treatment groups. Also, $\hat{b}$ indicates the estimate of the parameter $b$, while $\bar{b}$ indicates the sample statistics directly computed from the collected data. Accordingly, $g \in \{t,c\}$,

$$\bar{x}'_g = \sum_{i=1}^{N_g} x'_{g,i} / N_g, \bar{y}'_g = \sum_{i=1}^{N_g} y'_{g,i} / N_g, \bar{s}^2(x'_g) = \sum_{i=1}^{N_g} (x'_{g,i} - \bar{x}')^2 / N_g, \bar{s}^2(y'_g) = \sum_{i=1}^{N_g} (y'_{g,i} - \bar{y}'_g)^2 / N_g, \text{ and}$$

$$\bar{r}_g(x'_g, y'_g) = \sum_{i=1}^{N_g} (x'_{g,i} - \bar{x}'_g)(y'_{g,i} - \bar{y}'_g) / N_g \bar{s}(y'_g) \bar{s}(x'_g).$$

According to (3),

$$\hat{e}_o = \hat{m}(y_t) - \hat{m}(y_c) \qquad (11)$$

Where $\hat{m}(y_g)$ is the ML estimate of $\mathbf{E}[y_g]$.

The mean of the clinical endpoint over the pre-selection population, $\hat{m}(y_g)$, can be calculated from the estimates of the mean, $\hat{m}(x)$, and standard deviation, $\hat{\sigma}(x)$, of the selection measurement, $x$. They are provided in Cohen, 1955 as,

$$\hat{\sigma}(x) = v_1 / [S(\hat{z}) - \hat{z}], \qquad (12)$$

$$\hat{m}(x) = a - \hat{\sigma}(x)\hat{z}, \qquad (13)$$

where $\hat{z}$ is obtained by solving an estimating equation in (14),

$$[1-\hat{z}(S(\hat{z})-\hat{z})]/(S(\hat{z})-\hat{z})^2 - v_2/v_1^2 = 0, \tag{14}$$

where $v_k = \sum_{i=1}^{N_g}(x_i'-a)^k / N_g$, and $k \in \{1,2\}$.

Only when the distribution of the selection measurement, $x_i$, fairly closely follows a normal distribution, will Equation (14) have a valid root. This is one of major constraints that numerically prevent the Cohen ML method from being used in the cases where the data do not follow a normal distribution.

$\hat{m}(y_g)$ can then be estimated from $\hat{m}(x)$ and the data as

$$\hat{m}(y_g) = \hat{\alpha}_g + [\hat{m}(x) - \overline{x}_g']\hat{\beta}_g, \tag{15}$$

where $\hat{\alpha}_g = \overline{y}_g'$, $\hat{\beta}_g = \overline{r}_g(x_g', y_g')\overline{s}(y_g')/\overline{s}(x_g')$.

Since $\hat{\alpha}_g$, $\hat{\beta}_g$, $\hat{m}(x)$, and $\hat{m}(y_g)$ are all ML estimators, and, under the large sample assumption, they all approximate normal distributions. Cohen (1955) also shows that $\hat{\alpha}_g$, $\hat{\beta}_g$, and $\hat{m}(x)$ are uncorrelated with each other. If the number of patients for each treatment group, $N_g$, is pre-determined in the trial protocol, the asymptotic variances of these estimates are provided in Cohen (1955) as,

$$\mathbf{V}[\hat{\alpha}_g] = \sigma_g^2 / N_g \tag{16}$$

$$\mathbf{V}[\hat{\beta}_g] = \sigma_g^2 / [\sigma^2(x) N_g \phi_{11}(\hat{z})] \tag{17}$$

$$\mathbf{V}[\hat{m}(x)] = \sigma^2(x)\phi_{22}(\hat{z}) / [N_g(\phi_{11}(\hat{z})\phi_{22}(\hat{z}) - \phi_{12}(\hat{z})^2)] \tag{18}$$

where $\phi_{11}(\hat{z}) = 1 - S(\hat{z})[S(\hat{z}) - \hat{z}]$, $\phi_{12}(\hat{z}) = S(\hat{z})\{1 - \hat{z}[S(\hat{z}) - \hat{z}]\}$, and $\phi_{22}(\hat{z}) = 2 + \hat{z}\phi_{12}(\hat{z})$.

It is possible to derive the closed form of the variance of $\hat{m}(y_g)$, i.e. $\mathbf{V}[\hat{m}(y_g)]$, based on (15) and the uncorrelated relationship between $\hat{\alpha}_g$, $\hat{\beta}_g$, and $\hat{m}(x)$. However, the resultant form is too complicated to provide us with more insights, and thus omitted here. In fact, a straightforward way to calculate $\mathbf{V}[\hat{m}(y_g)]$ is using the Monte Carlo method. Since $\hat{\alpha}_g$, $\hat{\beta}_g$, and $\hat{m}(x)$ follow a univariate normal distribution with parameters estimable from (13)-(18), it is possible to generate samples of $\hat{m}(y_g)$ according to (15). $\mathbf{V}[\hat{m}(y_g)]$ can then be computed from the samples using the Monte Carlo method.

According to (11), the distribution of the estimate of the treatment effect, $\hat{e}_o$, should also be asymptotically normal. Its variance is

$$\mathbf{V}[\hat{e}_o] = \mathbf{V}[\hat{m}(y_t)] + \mathbf{V}[\hat{m}(y_c)], \tag{19}$$

since $\hat{m}(y_t)$ and $\hat{m}(y_c)$ are uncorrelated from each other.

3.2. Simulation

First, we would like to evaluate the bias and the statistical power of the Cohen ML method. The simulation settings are specified in Table 3.

Table 3

| Distribution | $\mu_c = \mu$ | $\mu_t$ | $\sigma_c = \sigma$ | $\sigma_t$ | $\rho_c$ | $\rho_t$ | $(a-\mu)/\sigma$ | $N_g$ |
|---|---|---|---|---|---|---|---|---|
| normal | 0.00 | 0.00, 0.15, 0.50 | 1.00 | 0.50, 0.85, 1.00, 1.15, 1.50 | 0.2 | 0.1, 0.2, 0.3, 0.6 | q(0.5), q(0.25) | 250, 500, 1000, 2000 |

The parameters in Table 3 can produce 480 parameter combinations, i.e. simulation scenarios. For each simulation scenario, 2000 trials are simulated. The bias is a function of the parameters in Table 3, and can be calculated as

$$Bias = \mathbf{E}[\hat{e}_o] - e_o = \sum_{i=1}^{2000} \hat{e}_o^{(i)} / 2000 - e_o,$$

where $\hat{e}_o^{(i)}$ is the estimated treatment effect over the pre-selection population in the $i^{th}$ trial using the Cohen ML method, and $e_o = (\mu_t - \mu_c)$ is the true treatment effect over the pre-selection population.

In order to provide a reference estimator for the Cohen ML method to compare with, a scenario suitable for using the conventional ANCOVA method was included. In this scenario, the patient selection procedure is removed. For each treatment group, $N_g$ patients are simulated, and they all go through the treatment stage.

Under the null hypothesis of no bias, i.e. $\mathbf{E}[\hat{e}_o] - e_o = 0$, $\mathbf{E}[\hat{e}_o]$ is claimed to be statistically significantly different from $e_o$ if the p-value of a t-test of $\hat{e}_o^{(i)} - e_o$, $i = 1 \cdots 2000$, is smaller than 0.05.

The simulation results about the estimation bias are reported in Figure 3. Like the simulation results presented in Subsection 2.2, the x-axis is defined as $\rho_t \sigma_t - \rho_c \sigma_c$. Our simulation shows that the biases are the same for three different choices of $\mu_t$, i.e. $\mu_t = 0.00$, 0.15, or 0.50. Therefore, only the results of the cases where $\mu_t = 0.00$ are presented in Figure 3.

Figure 3 has three panels. The first two panels present the simulation results obtained using Cohen ML method with two different selection thresholds, while the third panel presents the results obtained using the ANCOVA method to act as a reference. Each panel includes four groups of simulation results, and each group of results have the same number of patients, $N_g$. A smoothed curve is produced using the LOESS method for each group of results to highlight their general trends.

The results in Figure 3 show that the biases of the Cohen ML method are generally small and non-significant if the number of patients is sufficiently large, and the selection criterion is not very exclusive. The more selective the criterion is, the larger the bias is. Also, the bias is larger when the number of patients is relatively smaller. In addition, the simulation results suggest that the magnitude of the bias increases with the absolute values of $\rho_t \sigma_t - \rho_c \sigma_c$.

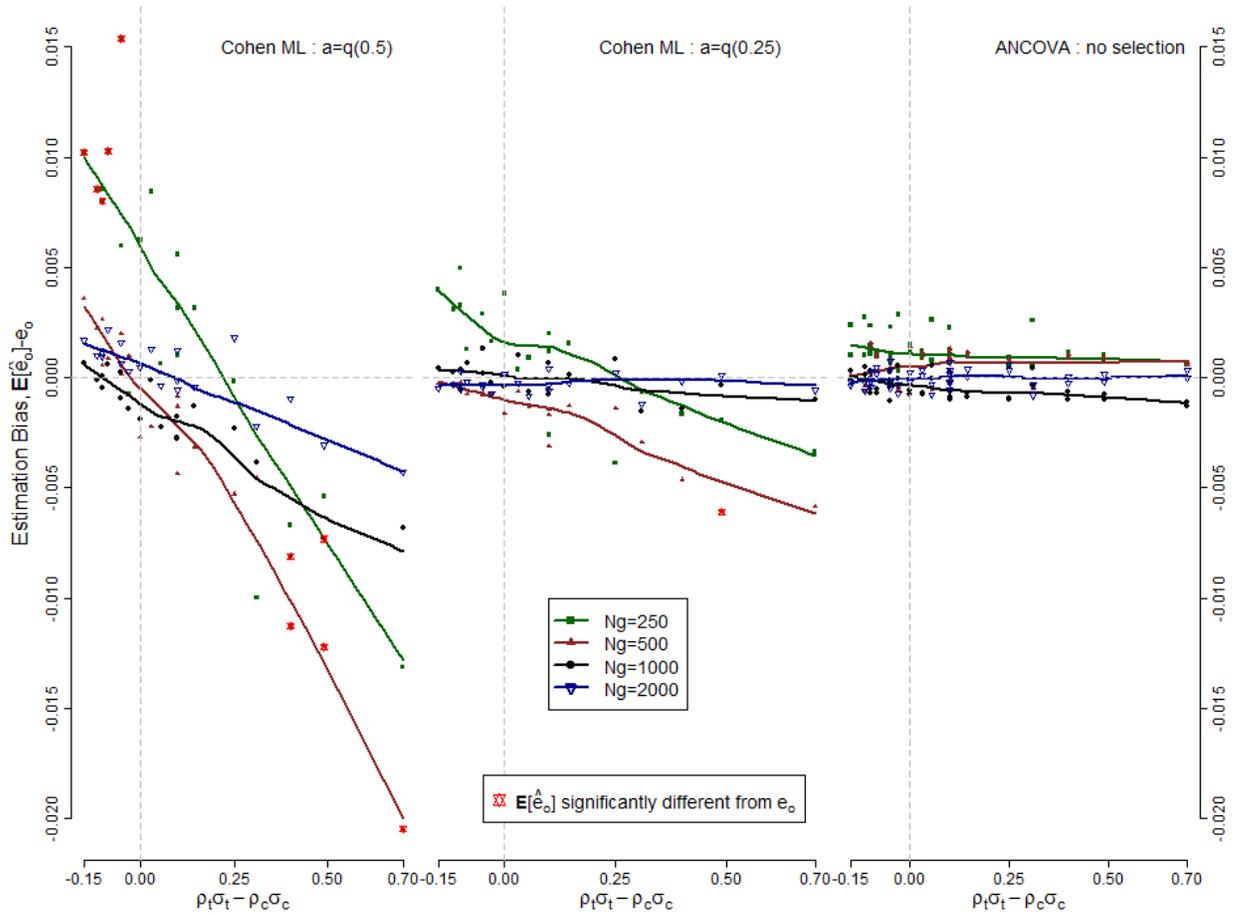

Figure 3 The bias of the Cohen ML method

In order to evaluate the statistical power of the Cohen ML method, we focus on the 180 simulation scenarios with $\mu_t = 0.15$. The statistical power is estimated as

$$Power = \sum_{i=1}^{2000} I_{0.05}(\hat{e}_o^{(i)}) / 2000,$$

where $I_{0.05}(\hat{e}_o^{(i)})$ is an indicator function, which is 1 when the estimate $\hat{e}_o^{(i)}$ is statistically significantly different from 0 at a significance level of 0.05, and 0 otherwise. The results are reported in Figure 4.

Figure 4 has 8 panels arranged in four rows and two columns. The two panels at each row have the same number of patients, while the four panels at each column have the same selection criterion. Within each panel, results with the same $\sigma_t$ value are plotted as a single curve.

Results in Figure 4 suggest that the statistical power of the Cohen ML method increases with a larger number of patients and decreases with a more selective criterion. Also, the power decreases with a larger variance of the clinical endpoints in the treatment group, $\sigma_t$, and increase with a smaller correlation between selection measurements and clinical endpoints, $\rho_t$. In a favorable scenario, e.g. $a = q(0.15), N_g = 2000$, the power of Cohen ML can approach 1.0. In contrast, in a less favorable scenarios, e.g. $a = q(0.5), N_g = 250$, the average power can be as low as about 0.1.

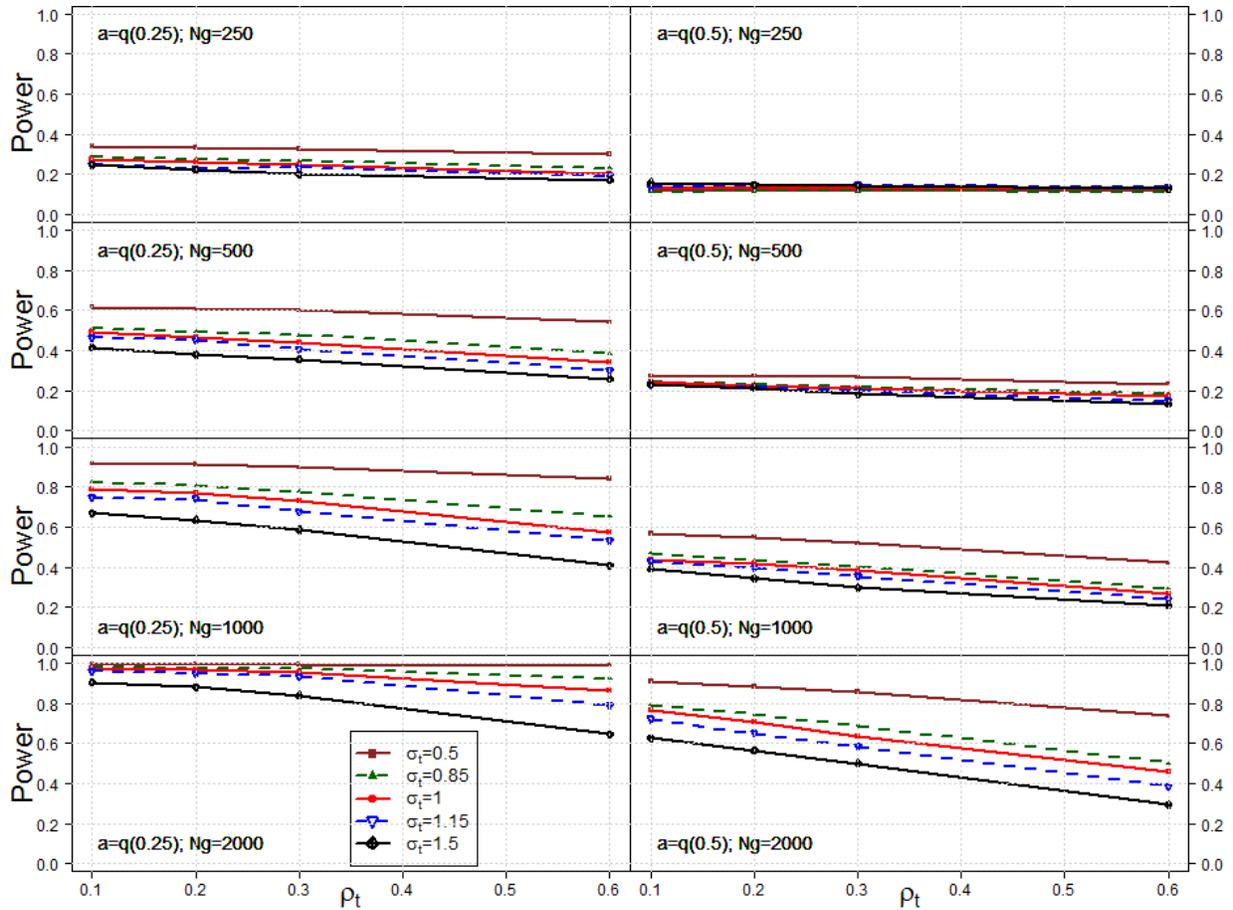

Figure 4 The power of the Cohen ML method

The simulations so far are based upon the assumption that the data strictly follow a bivariate normal distribution. To evaluate how much the Cohen ML method can tolerate a data distribution that deviates from this assumption, 120 simulation scenarios with the parameter combinations defined in Table 4 were used. A bivariate t-distribution with degrees of freedom of 8 is used to simulate 2000 trials for each simulation scenario. Since the bias is the same for the three choices of $\mu_t$, i.e. $\mu_t = 0.00$, 0.15, or 0.50, only the results with $\mu_t = 0.00$ are presented in Figure 5.

Table 4

| Distribution | $\mu_c = \mu$ | $\mu_t$ | $\sigma_c = \sigma$ | $\sigma_t$ | $\rho_c$ | $\rho_t$ | $(a-\mu)/\sigma$ | $N_g$ |
|---|---|---|---|---|---|---|---|---|
| $t(df = 8)$ | 0.00 | 0.00, 0.15, 0.50 | 1.00 | 0.50, 0.85, 1.00, 1.15, 1.50 | 0.2 | 0.1, 0.2, 0.3, 0.6 | q(0.5), q(0.25) | 1000 |

Results in Figure 5 show that the bias estimated using the Cohen ML method is large and is significantly different from 0 when $\rho_t\sigma_t - \rho_c\sigma_c \neq 0$. In contrast, the ANCOVA method tolerates the deviation in data distribution well, and its bias estimates remain close to 0. In fact, when a bivariate t-distribution with smaller degrees of freedom, e.g. 3 or 4, was used to simulate the data, the Cohen ML method in many cases numerically failed because Equation (14) did not have a valid root in those cases.

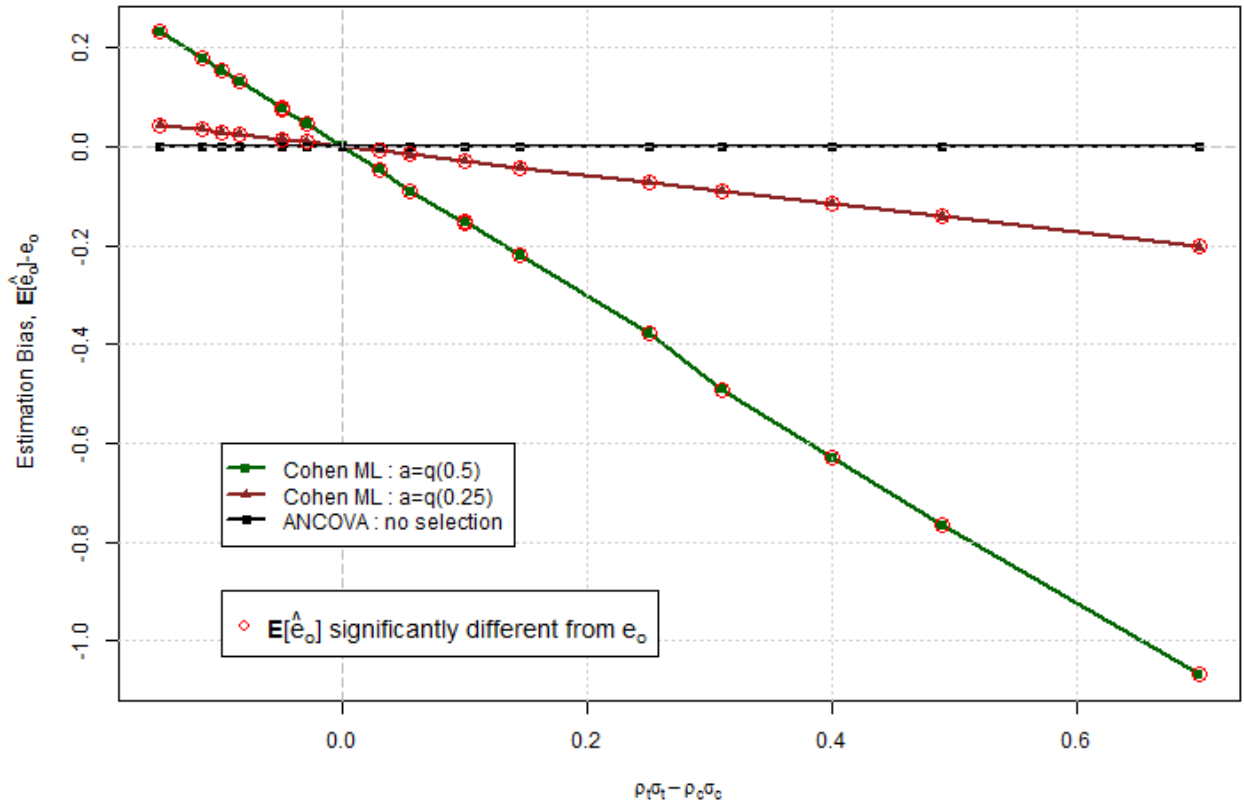

Figure 5 Large biases produced by the Cohen ML method with non-normal distributions

In summary, simulation results in this subsection suggest that, using existing Cohen ML method, it is hard to mitigate the SICE effect even under a very strong assumption on the data. It is somewhat possible only when the data are strictly normal, the selection criterion is mild, and the size of the trial is large.

## 4. APPLICATION TO A REAL DATASET

In this section, a clinical trial dataset from a real drug development program is used to illustrate the impact of the SICE effect. The involved compound, coded as MK in this paper, was developed until 2007 as a treatment for insomnia.

This dataset was from a placebo-controlled randomized parallel trial, which included both a treatment arm of MK and a placebo arm. Each patient slept at a sleep center on two separate visits to measure their objective sleep quality, including Latency to Persistent Sleep (LPS). The first visit happened before the patient took any treatment and was used as the baseline night. During the second visit, the patient first took either MK or placebo according to the treatment group he/she was randomized to MK or placebo, and then his/her sleep quality was measured for the second time. As a treatment for insomnia, MK was expected to make patients to fall asleep faster. That is, compared with the patients in the placebo group, those in the MK group were expected to have smaller LPS.

It is a fairly common practice in insomnia clinical trials that patients are evaluated and selected at the baseline to make sure they are indeed primary insomnia patients. One of the frequently used criteria is to select only patients whose LPS exceeds a threshold, say T. That is, that a patient needs at least T minutes to enter stable sleep. However, there is no agreement between practitioners on the exact value of T. Therefore, some trials use 15 minutes, while the others 20 minutes, as the threshold. It is subsequently demonstrated that, due to the SICE effect, different choices of the threshold T can have a dramatic impact on the size of the observed drug effect.

In order to make the LPS data more normal, a log-transform is first applied to them. The drug effects are estimated using both the t-test and the ANCOVA method. The drug effect over the pre-selection population is first estimated. After that, a selection procedure is introduced at the baseline with 6 different selection thresholds ranging from 15 minutes to 20 minutes, i.e. T= 15, 16, …, 20. The drug effect under each threshold is then estimated based upon the corresponding selected patients. The results obtained using both t-test and ANCOVA methods are reported in Figure 6. The point estimate is marked, along with its 95% confidence interval. The number of patients in each treatment group under each selection threshold is shown at the top of the figure.

Figure 6 shows that, due to the SICE effect, the observed MK drug effect is generally increased with more selective thresholds. The observed MK's effect on LPS when T=15 minutes is 39% larger than that in the scenario without patient selection. In contrast, the observed MK's effect when T=20 minutes is 106% larger. The concept of the SICE effect helps us understand that the selection threshold T plays such an important role in the size of the observed drug effect. In other words, the observed drug effect can be very sensitive to the choice of selection threshold in this dataset.

Based on this observation, we recommend that, when analyzing a large confirmatory clinical trial, if possible, we should repeat the analysis several times with different selection thresholds, or adjusted versions, of each inclusion/exclusion criterion to evaluate how sensitive the observed treatment effect is to this particular selection criterion. For example, if this trial originally used a threshold of 15 minutes, and thus it only collected subjects whose baseline LPS is no fewer than 15 minutes, it is still possible to implement our recommendation by increasing the baseline LPS thresholds from 16 to 20 minutes using the available data.

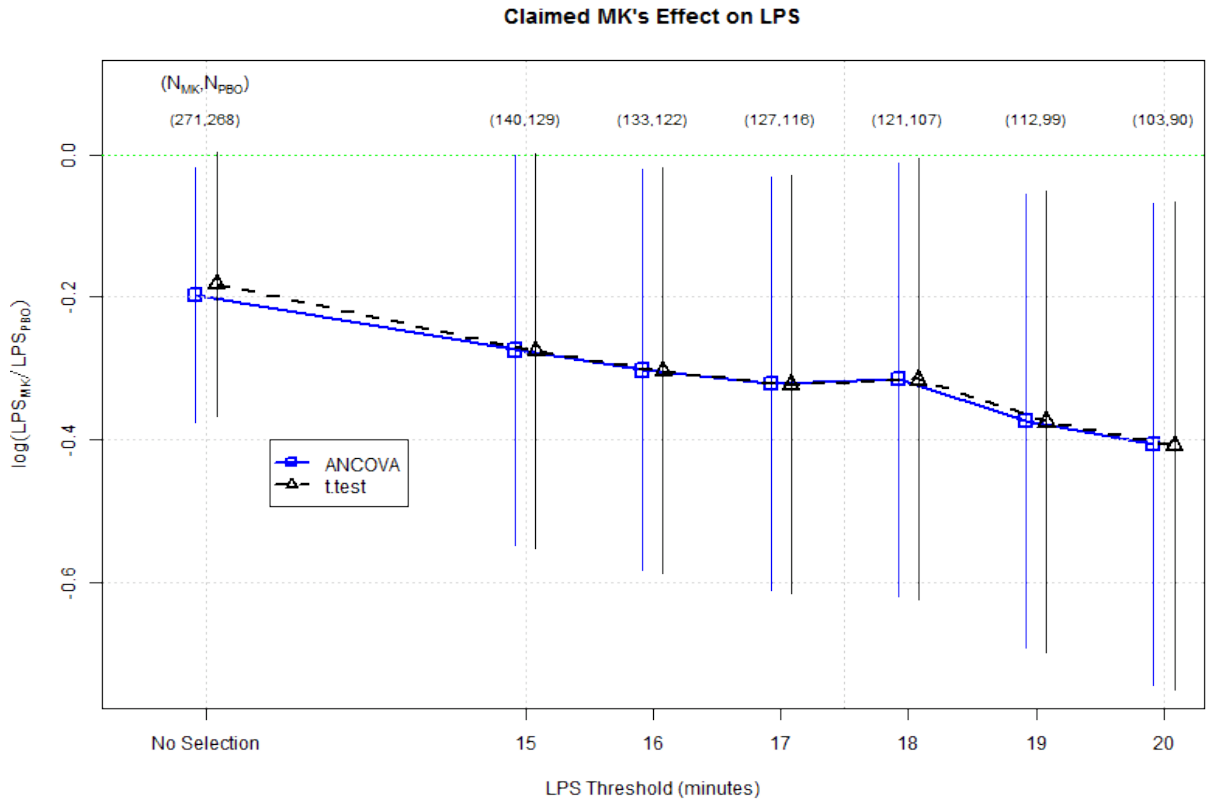

Figure 6 Observed MK's effect on LPS with or without patient selection

Since the log-transformed LPS data approximately follow a normal distribution, the Cohen ML method is applied to estimate the MK's effect over the pre-selection population at each chosen LPS threshold. The estimates, along with their 95% confidence intervals, are reported in Figure 7. The drug effect directly estimated with the pre-selection population is also reported in Figure 7 for comparison. Figure 7 confirms our observation in Section3, and suggests that, compared with the estimate obtained with the original dataset, the estimates obtained from the selected populations have relatively large biases and wide 95% confidence intervals, and are thus not very informative.

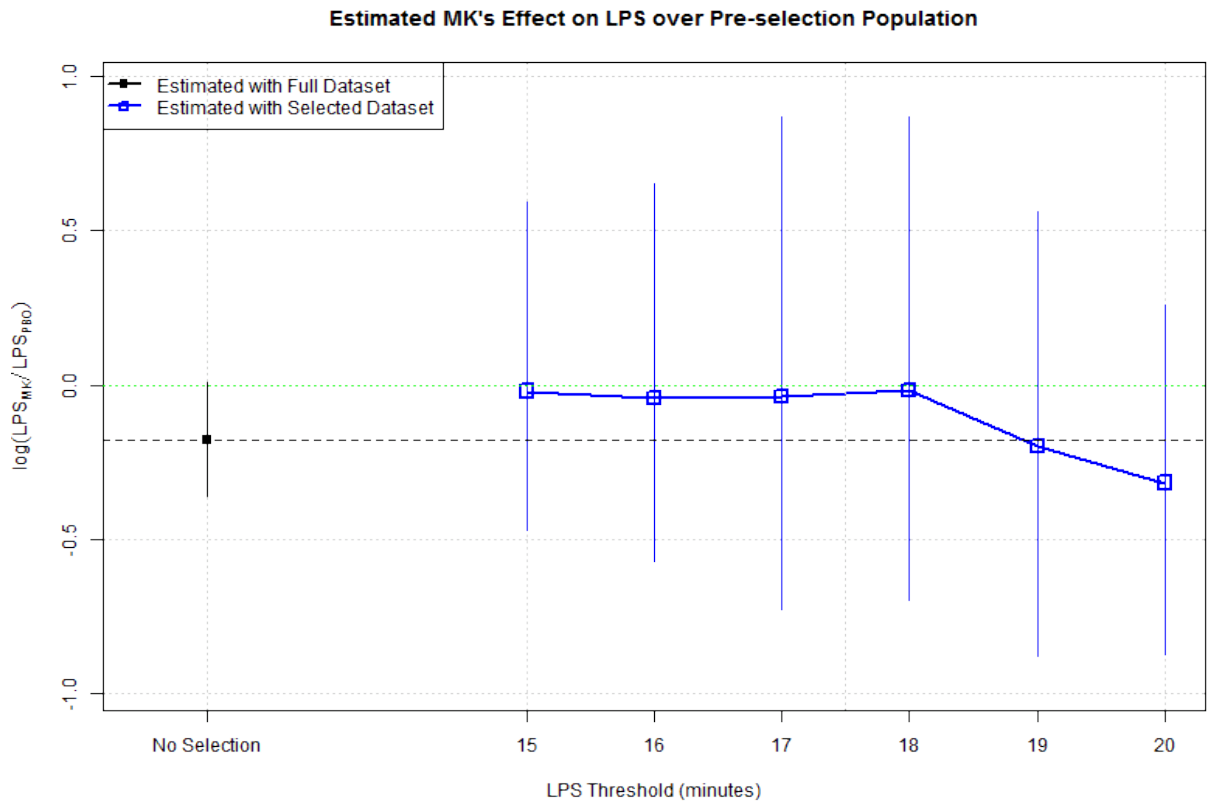

Figure 7 MK's effect over pre-selection population estimated using Cohen ML method

## 5. SUMMARY AND DISCUSSION

In order to better understand the impact of a patient selection criterion on the treatment effect from a RCT, a concept, called the SICE effect, is introduced and formulated in this paper. The effect occurs when the covariance of the measurements with a variable used for selection are different between treatment groups. Since this effect is proportional to the difference between two covariance estimates, its magnitude is usually not large, and its impact is expected to become evident primarily in large-scale confirmatory clinical trials.

One of our major motivations to study the SICE effect is to advocate for a more transparent understanding of the I/E criteria in RTCs. Both the simulation results and the results from a clinical trial dataset demonstrate that, due to the SICE effect, a selection criterion can have a surprisingly large impact on the magnitude of the observed treatment effect. Based upon this observation, when possible, the sensitivity of an observed treatment effect on a selection criterion should be evaluated by repeating the data analysis with several adjusted versions of the selection criterion, e.g. using different thresholds in the selection criterion. Those I/E items that can substantially impact the observed treatment effects should be duly noted.

The SICE effect is studied in this paper under an assumption that the measurements are numeric variables, which follow bivariate normal distributions. However, the concept of the SICE effect exists in clinical trials with other types of clinical outcomes, such as probability for an event to happen, time to event, etc. Insights gained in this paper could be expanded to other types of clinical outcomes.

The essence of the SICE effect has implications on other statistical areas, such as missing data analysis. For example, in a large trial studying a pain treatment, those patients suffering more severe pains have a higher probability to drop out. The effect of drop-outs can be viewed as a patient selection procedure with a soft/fuzzy threshold. When analyzing this type of data, besides considering other known effects, it may be beneficial to also pay attention to the SICE effect induced by the drop-outs.

## APPENDIX

**Appendix A**: R code of the simulated simple two-treatment parallel trial

```
library(MASS)
# To set up the parameters of two bivariate norm distributions
M1 = array(0,c(2,1));
Sigma1 = matrix(c(1,0.2,0.2,1), ncol=2);
```

```
M2 = array(0,c(2,1));
Sigma2 = matrix(c(1,0.3*1.2,0.3*1.2,1.2^2), ncol=2);
# To set up the threshold
a = 0.6;
# To define the number of patients planning to recruit for each treatment group.
nSubj = 1000;
set.seed(500);
# To generate the pre-selection population
nTotalSubj = round(nSubj/(1-pnorm(a, mean = 0, sd = 1)));
PBO0 = mvrnorm(nTotalSubj, M1, Sigma1);
DRUG0 = mvrnorm(nTotalSubj, M2, Sigma2);
# The selection procedure
PBO = PBO0[PBO0[,1]>a, 2]
Drug = DRUG0[DRUG0[,1]>a, 2]
# estimate treatment effect using a 2-sample t.test
tRes = t.test(Drug, PBO)
```

**Appendix B**: Derivations of the results in Section 2.1.

Let two i.i.d. random variables $z_1, z_{g,2} \sim N(0,1)$, and $g \in \{t,c\}$. Using Cholesky decomposition, (6) can be reformulated as

$$\begin{bmatrix} x \\ y_g \end{bmatrix} = \begin{bmatrix} \mu \\ \mu_g \end{bmatrix} + \begin{bmatrix} \sigma & 0 \\ \rho_g \sigma_g & \sigma_g\sqrt{1-\rho_g^2} \end{bmatrix} \begin{bmatrix} z_1 \\ z_{g,2} \end{bmatrix},$$

which is equivalent to

$$\begin{aligned} x &= \mu + \sigma z_1 \\ y_g &= \mu_g + \rho_g \sigma_g z_1 + \sigma_g \sqrt{1-\rho_g^2} z_{g,2} \end{aligned} \qquad (20)$$

Equation (20) implies that x and $z_{g,2}$ are independent from each other, and thus $\mathbf{E}[z_{g,2} \mid x > a] = 0$.

Therefore, $\mathbf{E}[y_g \mid x > a] = \mu_g + \rho_g \sigma_g \mathbf{E}[z_1 \mid x > a]$, (21)

where $\mathbf{E}[z_1 \mid x > a] = \mathbf{E}[z_1 \mid z_1 > \frac{a-\mu}{\sigma}] = S(\frac{a-\mu}{\sigma})$, according to (20).

Bringing (21) into (4), and then into (5) along with (3), (7) can be obtained. □

To avoid the tedious derivation of the conditional variance of $y_g$, $\mathbf{V}[y_g \mid x > a]$, we directly cite its form from Heckman (1979) as

$$\mathbf{V}[y_g \mid x > a] = \sigma_g^2 - \rho_g^2 \sigma_g^2 S(z)(S(z) - z). \quad (22)$$

Since $y_c'$ and $y_t'$ are two independent samples, according to (8),

$$\mathbf{V}[\hat{e}_s] = \mathbf{V}[\bar{y}_t'] + \mathbf{V}[\bar{y}_c'] = \mathbf{V}[y_t \mid x > a] / N_t + \mathbf{V}[y_c \mid x > a] / N_c. \quad (23)$$

Equation (9b) can be obtained by bringing (22) into (23). □

Bringing the first equation in (20) into (10), it can be obtained that

$$\frac{\mathbf{E}[y_g \mid x > a] - \mu_g}{\sigma_g} = \rho_g \mathbf{E}[\frac{x-\mu}{\sigma} \mid x > a] = \rho_g \mathbf{E}[z_1 \mid x > a],$$

which is exactly the same as (21). Thus, the difference between (10b) and (10a) is exactly (7). □